# Financial Time Series: Stylized Facts for the Mexican Stock Exchange Index Compared to Developed Markets'

Omar Rojas-Altamirano (orojas@up.edu.mx)

Carlos Trejo-Pech (ctrejo@up.edu.mx)

School of Business and Economics

Universidad Panamericana Campus Guadalajara, México

# Content

## List of Figures

## List of Tables

**Abstract**

In this chapter we present some stylized facts exhibited by the time series of returns of the Mexican Stock Exchange Index (IPC) and compare them to a sample of both developed (USA, UK and Japan) and emerging markets (Brazil and India). The period of study is 1997-2011. The stylized facts are related mostly to the probability distribution function and the autocorrelation function (e.g. fat tails, non-normality, volatility clustering, among others). We find that positive skewness for returns in Mexico and Brazil, but not in the rest, suggest investment opportunities. Evidence of nonlinearity is also documented.

**Introduction**

It is widely agreed on that Mathematical Finance, and in particular the study of financial time series from a statistical point of view, started on March 29$^{th}$, 1900 at La Sorbonne, Paris, when Louis Bachelier, Poincaré's student, presented his thesis *Théorie de la Spéculation,* c.f. (Bachelier, 1900). Bachelier introduced the theory of Brownian motion used for the modeling of price movements and the evaluation of contingent claims in financial markets (Courtault et al., 2000). In Bachelier's proposition, if $P_t$ denotes the price of an asset, at some period *t*, then $P_{t+1} = P_t + \varepsilon_t$ is the price of the asset at a future unit instance, assuming that $\varepsilon_t \sim N(\mu, \sigma^2)$. Such an asumption was, in fact, a preamble of the Efficient Market Hypothesis developed much later by Fama (1965). However, hidden in the chaos of this pure stochastic process, some patterns in the price movement were recognized by Osborne (1959). In particular, Osborne showed that from moment to moment the market is much more likely to reverse itself than to continue on a trend. However, when price moved in the same direction twice, it was much more likely to continue in that direction than if it had moved in a given direction only once (Weatherall, 2013). Other behaviours, common to a wide variety of assets, have also been documented. For instance, the non-Gaussianity of returns -contradicting Bachelier's original assumption, as empirical distributions of price changes are usually too peaked to be relative to samples from Gaussian populations (Mandelbrot, 1963). From then on, sets of properties, common across many instruments, markets and time periods, have been observed and termed *stylized facts* (Cont, 2001).

Some of these stylized facts relate to the shape of the probability distribution function of returns. In particular, empirical studies report that the distributions are *leptokurtic* (more peaked and with fatter tails than those corresponding to the normal distribution) and *skewed*. Using a kernel density estimator, it can be shown that most distributions of returns can be adjusted by a fat-tail distribution, e.g. a Student's *t*-distribution with 3 to 5 degrees of freedom. On the other hand, as one increases the time-scale on which returns are measured, the distributions tend to Gaussianity. Another set of stylized facts can be derived from the *autocorrelation function* (ACF). For instance, a slow decay of the ACF in absolute returns, the absence of correlation after some time, and the formation of volatility clusters. Some of these stylized facts have been empirically confirmed on some indexes and exchange rates, see, e.g. (Franses & van Dijk, 2000) and (Zumbach, 2013). The case of the Belgrade Stock Market was treated in (Miljković & Radović, 2006). For a more exhaustive list of references, see (Sewell, 2011).

After observing the predominant stylized facts of returns of a given asset, and running some normality and linearity tests, could be in a more comfortable position to chose a proper model, most of the times non-linear, for the empirical financial data at hand.

We study some of these stylized facts exhibited by the Mexican Stock Exchange Index (IPC), and compare them to indexes from both developed and emerging markets (USA [S&P 500], UK [FTSE], JAPAN [NIKKEI 225], Brazil [IBOVESPA], and India [BSE]). Some authors have already focused their attention on the IPC, studying different, but related, empirical problems. Asymmetric ARCH models where used to model the daily returns of 30 stocks of the IPC by (Lorenzo Valdéz & Ruíz Porras, 2011), and by (López Herrera, 2004). These studies focused on volatility of returns. We believe this chapter to be of interest, since it provides the stylized facts on the IPC returns time series. The knowledge of those facts could be helpful to determine better empirical models, most of the times nonlinear, to produce reliable forecasts.

**Stylized Facts of Returns for the Mexican Stock Exchange Index**

In this section we present some of the *stylized facts* exhibited by the Mexican Stock Exchange Index (IPC) and compare them to those from other developed and emerging markets (USA [S&P 500], UK (FTSE), JAPAN [NIKKEI 225], Brazil [IBOVESPA], and India [BSE]). Daily-adjusted closing prices from January 1997 to December 2011 are used, see Figure 1.

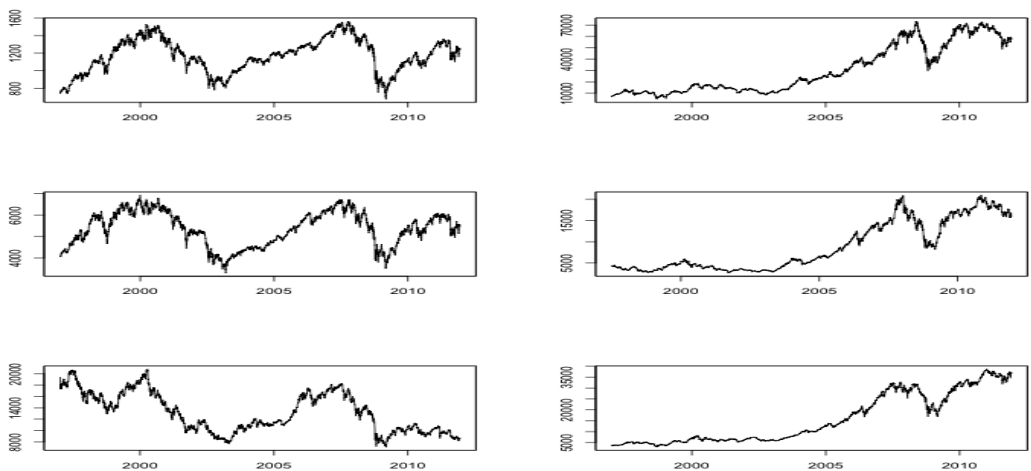

**Figure 1 - Daily observations on the level of the stock indexes of developed markets (left), from top to bottom USA (S&P 500), United Kingdom and Japan, and emerging markets (right), from top to bottom Brazil, India and Mexico from January 1997 to December 2011.**

As can be observed from Figure 1, the time series of the stock indexes do not seem to have anything in common. However, as we will see in the next section, returns do.

**From prices to returns**

Most financial studies involve returns of assets instead of prices. According to Campbell et al. (1997), there are two main reasons for using returns. First, for average investors, returns represent a complete and scale-free summary of the investment opportunity. Second, return series are easier to handle than price series because the former have more attractive statistical properties. There are, however, several definitions of an asset return.

Let $P_t$ be the price of an asset at time $t$. The *simple return* of an asset of price $P_t$ is given by

$$R_t = \frac{P_t - P_{t-1}}{P_{t-1}},$$

from where

$$1 + R_t = \frac{P_t}{P_{t-1}}.$$

The natural logarithm of the above *gross return* in percentual terms, leads to the *continuously compounded percentual return*

$$r_t = 100 \cdot (p_t - p_{t-1}),$$

where $p_t = \ln(P_t)$ and $r_t = \ln(1 + R_t)$. We will focus our attention throughout the rest of this chapter, on the time series of returns (also called *percentual log-returns*) defined by $\{r_t\}$. Figure 2 plots the time series of returns for the different indexes under study.

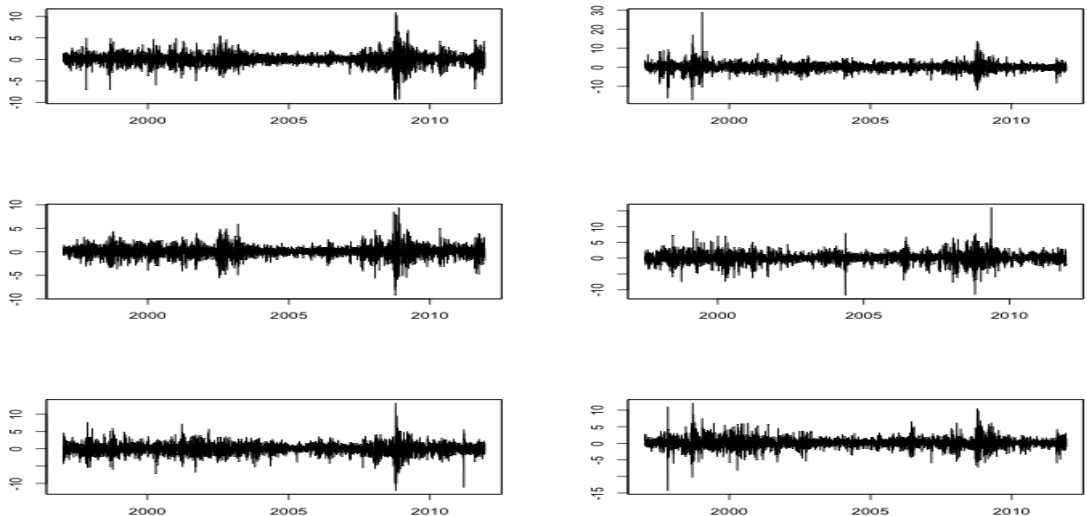

**Figure 2 - Time series of returns corresponding to stock indexes of developed markets (left), from top to bottom USA (S&P 500), United Kingdom and Japan, and emerging markets (right), from top to bottom Brazil, India and Mexico from January 1997 to December 2011.**

**Probability Density Function**

A traditional assumption in financial mathematics, convenient to make statistical properties of returns tractable, is that $r_t : i.i.d. \ N(\mu, \sigma^2)$. However, as it has been known since (Mandelbrot, 1963), such assumption encounters difficulties when empirically tested. To illustrate, cf. to Figure 3, where the peak of the histogram is much higher than the corresponding to the normal distribution and it is slightly skewed to the right.

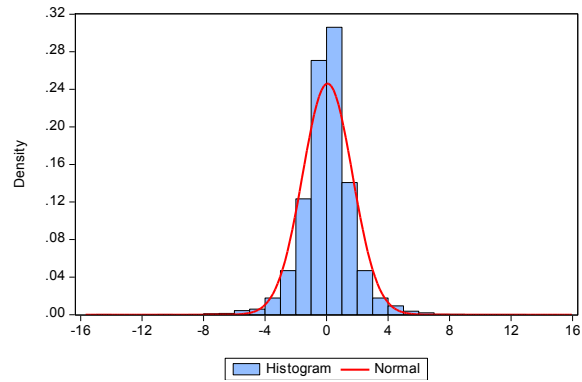

Figure 3 - Histogram of daily returns of the IPC against the theoretical normal distribution

Summary statistics for daily index returns $r_t$ from 1997 to 2011 are provided in Table 1. These statistics are used in the discussion of some stylized facts related to the probability density function of the series below.

**Table 1 - Summary statistics for stock index returns**

| Index | Mean | Median | Min | Max | StdDev | Skewness | Kurtosis |
|---|---|---|---|---|---|---|---|
| **S&P 500** | 0.0137 | 0.0687 | -9.4695 | 10.9571 | 1.3501 | -0.2040 | 9.7826 |
| **FTSE** | 0.0077 | 0.0413 | -9.2645 | 9.3842 | 1.2918 | -0.1203 | 8.0672 |
| **NIKKEI** | -0.0220 | 0.0037 | -12.1110 | 13.2345 | 1.6048 | -0.2861 | 8.5632 |
| **IBOVESPA** | 0.0570 | 0.1379 | -17.2082 | 28.8324 | 2.2520 | 0.3184 | 15.3954 |
| **BSE** | 0.0365 | 0.1062 | -11.8091 | 15.9899 | 1.7193 | -0.0899 | 8.1902 |
| **IPC** | 0.0642 | 0.1073 | -14.3144 | 12.1536 | 1.5955 | 0.0131 | 9.4692 |

**Gain/loss asymmetry**

The *skewness* $\hat{S}$ of $r_t$ is a measure of the asymmetry of the distribution of $r_t$. The sample skewness can be estimated consistently by

$$\hat{S} = \frac{1}{n} \sum_{t=1}^{m} \frac{(r_t - \hat{\mu})^3}{\hat{\sigma}^3}.$$

Remember that all symmetric distributions, including the normal distribution, have skewness equal to zero. With the exception of Mexico and Brazil, most indexes returns have negative skewness (Table 1). This might point into possible opportunities of investment in these developing markets, since negative (positive) skewness implies that the left (right) tail of the distribution is fatter than the right (left) tail, or that negative (positive) returns tend to occur more often than large positive (negative) returns (Franses & van Dijk, 2000).

**Fat tails**

A random variable is said to have *fat tails* if it exhibits more extreme outcomes than a normally distributed random variable with the same mean and variance (Danielsson, 2011). This implies that the market has more relatively large and small outcomes than one would expect under the normal distribution.

The *kurtosis* measures the degree of peakedness of a distribution relative to its tails. The sample kurtosis can be estimated by

$$\hat{K} = \frac{1}{n} \sum_{t=1}^{n} \frac{(r_t - \hat{\mu})^4}{\hat{\sigma}^4}.$$

High kurtosis generally means that most of the variance is due to infrequent extreme deviations than predicted by the normal distribution that has kurtosis equal to 3. Such leptokurtosis is a signal of fat tails. As seen in Table 1, all stock index returns have excess kurtosis, well above 3, which is evidence against normality.

The most commonly used graphical method for analyzing the tails of a distribution is the quantile-quantile (QQ) plot. QQ plots are used to assess whether a set of observations has a particular distribution. The QQ plots for the IPC returns against some theoretical distributions are shown in Figure 4.

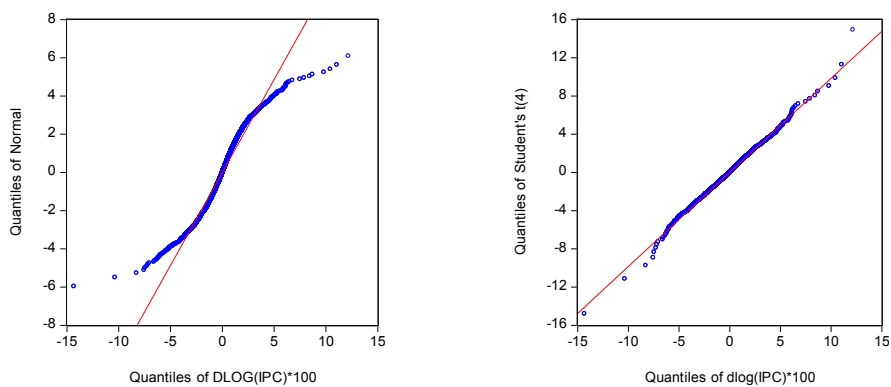

**Figure 4 - QQ plot of IPC returns against Normal (left) and Student t-distribution with 4 degrees of freedom (right)**

Returns seem to have fatter tails to fit the normal distribution. To have a sense on how fat the tails are, the Student t-distribution is used as this is a distribution with fat tails, where the degrees of freedom indicate how fat the tails actually are. Figure 5 shows an almost perfect fit by the Student t-distribution to the kernel density of IPC returns.

The fact that the distribution of returns is fat-tailed has important financial implications, especially because it leads to a gross underestimation of risk, since the probability of observing extreme values is higher for fat-tail distributions compared to normal distributions. Alan Greenspan (1997) warned financial markets on this: "The biggest problems we now have with the whole evolution of risk is the fat-tail problem, which is really creating very large conceptual difficulties. Because as we all know, the assumption of normality enables us to drop off the huge amount of complexity in our equations. Because once you start putting in non- normality assumptions, which is unfortunately what characterizes the real world, then these issues become extremely difficult".

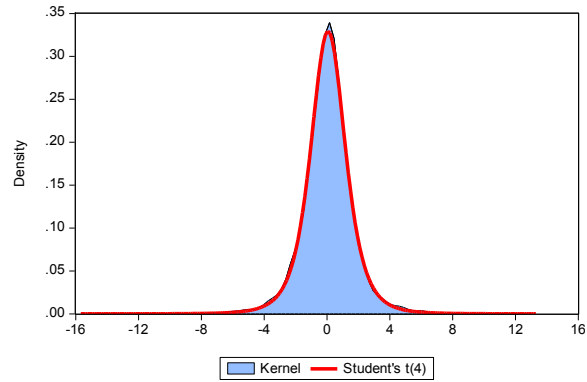

Figure 5 - Kernel density of IPC returns against Student t-distribution with four degrees of freedom

**Normality tests**

Two of the most common tests for normality are the Kolmogorov-Smirnov and the Jarque-Bera. The Jarque-Bera (JB) test is given by the statistic

$$JB = \frac{\hat{S}^2}{6/T} + \frac{(\hat{K}-3)^2}{24/T},$$

which is asymptotically distributed as a $\chi^2$ random variable with 2 degrees of freedom, where $\hat{S}$ is the sample skewness, $\hat{K}$ the sample kurtosis and $T$ the sample size. One rejects $H_0$ of normality if the *p*-value of the JB statistic is less than the significance level (Jarque & Bera, 1987). JB statistics for all 6 stock index return series are: S&P 500 (7237.259), FTSE (4046.752), NIKKEI (4778.961), IBOVESPA (23730.45), BSE (4010.862) and IPC (6532.395). The *p*-values = 0.0000, for all, reject normality.

## Agregational Gaussianity

As one increases the time scale over which returns are calculated, their distribution looks more and more like a normal distribution. In particular, the shape of the distribution is not the same at different time scales (Cont, 2001). Table 2 shows how the kurtosis and the value of the JB statistic decrease as the time scale increases. Daily, weekly and monthly returns all have a JB *p*-value that rejects normality; however, quarterly returns do not, as shown in Figure 6. In empirical research quarterly returns are seldom used.

**Table 2 - Summary statistics for IPC returns taken at different time scales from 1991 to 2011**

| Time scale | Mean | Median | StdDev | Skewness | Kurtosis | JB | JB *p*-value |
|---|---|---|---|---|---|---|---|
| **Daily** | 0.0648 | 0.0800 | 1.6201 | 0.0201 | 8.3821 | 6075.06 | 0.0000 |
| **Weekly** | 0.3107 | 0.5727 | 3.6367 | -0.2313 | 5.7106 | 330.83 | 0.0000 |
| **Monthly** | 1.3642 | 2.2857 | 7.6775 | -0.8049 | 5.2890 | 78.6402 | 0.0000 |
| **Quarterly** | 4.0528 | 4.7035 | 12.9816 | -0.0250 | 2.5772 | 0.6417 | 0.7255 |

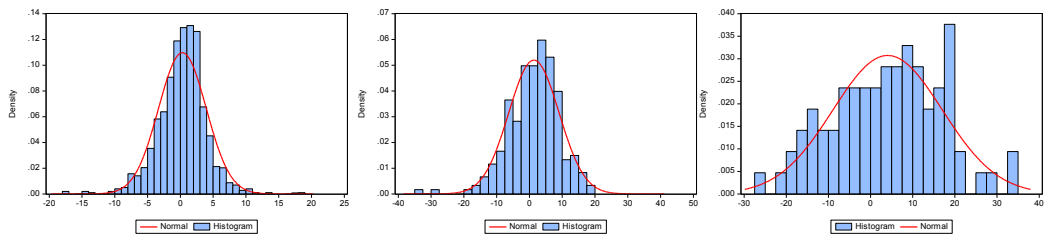

**Figure 6 - Histogram vs. theoretical normal distribution for IPC returns. Weekly (left), monthly (center) and quarterly (right)**

**Autocorrelation Function**

The lag-k *autocorrelation function* (ACF) of a time series $r_t$ is defined by

$$\rho_k = \frac{\gamma_k}{\gamma_0} = \frac{Cov(r_t, r_{t-k})}{Var(r_t)}.$$

The ACF measures how returns on a given day are correlated with returns on previous days. If such correlations are statistically significant, we have strong evidence for predictability.

**Absence of linear autocorrelation**

It is a well-known fact that price movements in liquid markets do not exhibit any significant linear autocorrelation (Cont, 2001). It is seen in Figure 7 how the autocorrelation function for the IPC series rapidly decays to zero after a lag (day). For more on the absence of significant linear autocorrelations in asset returns, cf. (Fama, 1965).

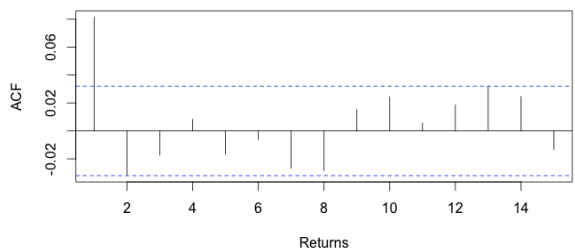

Figure 7 - Autocorrelation plot of IPC returns, along with a 95% confidence interval, for the first 15 lags

**Volatility clusters**

The most common measure of market uncertainty is volatility (the standard deviation of returns). A standard graphical method for exploring predictability in statistical data is the ACF plot. Figure 8, top panel, shows the ACF plot of IPC returns, along with a 95% confidence interval, from where it is evident that most autocorrelations lie within the interval. In contrast, Figure 8, middle panel and bottom panel, show the ACF plot of squared and absolute returns, respectively, where the ACF is significant even at long lags, providing strong evidence for the predictability of volatility, given the persistence of the autocorrelations. For various indices and stocks, it has been shown that the squared ACF of returns remains positive and decays slowly, remaining significantly positive over several days. This phenomenon is what is usually called the autoregressive conditional heteroscedasticity (ARCH) effect (Engle, ARCH: Selected Readings, 1995).

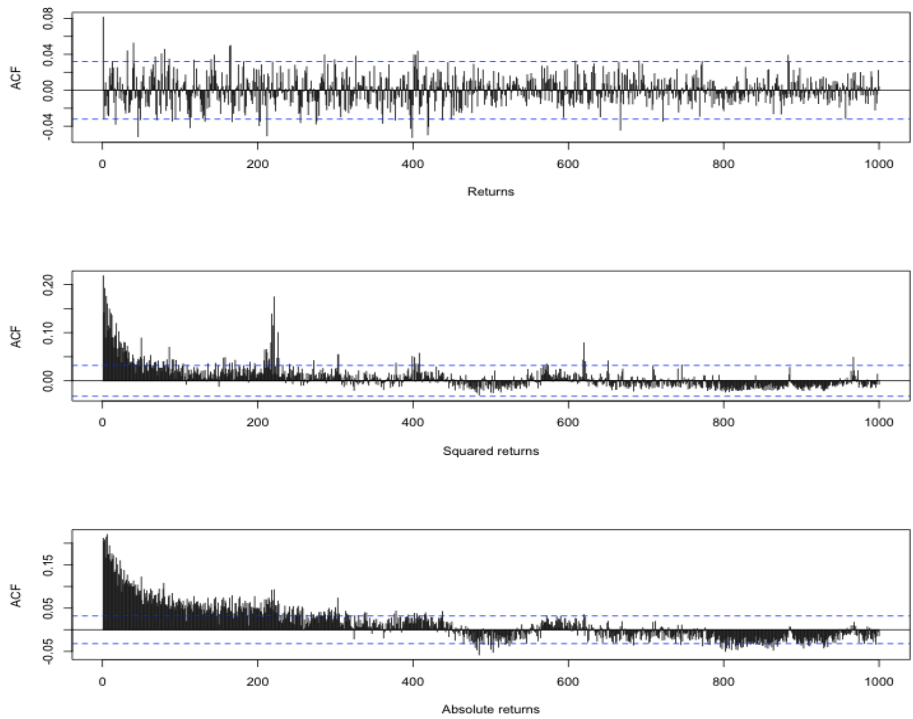

**Figure 8 - Autocorrelation plots of daily IPC returns 1997-2011 (top), squared returns (middle) and absolute returns (bottom). All the plots with a 95% confidence interval**

The *Ljung-Box* (LB) test (Ljung & Box, 1978) is used to test the joint significance of autocorrelation coefficients over several lags. It is a Portmanteau statistical test for the null hypothesis $H_0 : \rho_1 = \text{L} = \rho_m = 0$ against the alternative hypothesis $H_a : \rho_i \neq 0$ for some $i \in \{1, \text{K}, m\}$. It is given by

$$LB(m) = T(T+2) \sum_{l=1}^{m} \frac{\hat{\rho}_l^2}{T-l} .$$

The decision rule is to reject $H_0$ if the *p*-value is less than or equal to the significance level.

We used the LB test using 21 lags (approximately the number of trading days on a given month) of daily IPC returns. We tested using the full sample size (3,746 observations), as well as the most recent 1,000 and 100 observations. We performed the test on returns, square returns and absolute returns.

Table 3 - Ljung-Box test for daily IPC returns, squared returns and absolute returns, using 21 lags

| Time series | Sample size | Ljung-Box test | *p*-value |
| --- | --- | --- | --- |
| **IPC returns** | 3746 | 56.4024 | 4.403e-05 |
| **IPC returns** | 1000 | 51.0081 | 0.0002 |
| **IPC returns** | 100 | 29.1682 | 0.11 |
| **IPC squared returns** | 3746 | 1254.434 | 2.2e-16 |
| **IPC squared returns** | 1000 | 962.7398 | 2.2e-16 |
| **IPC squared returns** | 100 | 19.9187 | 0.5264 |
| **IPC absolute returns** | 3746 | 2317.262 | 2.2e-16 |
| **IPC absolute returns** | 1000 | 1268.726 | 2.2e-16 |
| **IPC absolute returns** | 100 | 30.6823 | 0.07911 |

Table 3 shows that there is significant return predictability for the full sample and the last 1,000 observations using returns, square returns and absolute returns. Using the last 100 observations, the data are independently distributed, i.e., no correlations amongst the observations. This does not imply a violation of market efficiency, since we would need to consider the risk free rate, adjust returns for risk, and include transaction costs (Danielsson, 2011). It is also shown how *p*-values for square and absolute returns are much smaller than for returns, suggesting how nonlinear functions of returns show significant positive autocorrelation or persistence. This is a quantitative sign of the stylized fact known as *volatility clustering*: large price variations are more likely to be followed by large price variations. Thus, returns are not random walks (Campbell, Lo, & MacKinlay, 1997).

**Volatility/return clusters**

Another way to possibly characterize volatility and return clusters is by looking at lag plots of returns, i.e., scatterplots of $r_t$ against $r_{t-1}$. A stylized fact that can be observed from such plots is that large returns tend to occur in clusters, i.e., it appears that relatively volatile periods characterized by large returns alternate with more stable periods in which returns remain small. Figure 9 shows the lag plots corresponding to returns of the S&P 500, the IBOVESPA and the IPC index. From these plots, it is apparent the aforementioned stylized fact.

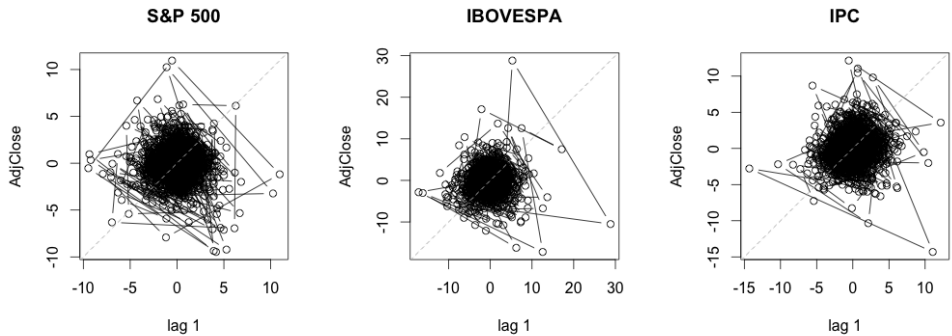

**Figure 9 - Lag plots of the returns on the S&P 500 (left), IBOVESPA (center) and IPC (right), on day *t*, against the return on day *t-1***

In order to concentrate on a partial route followed by the IPC return series, in Figure 10 we focus our attention on what appears to be the most volatile section of the lag plot from Figure 9 (right panel).

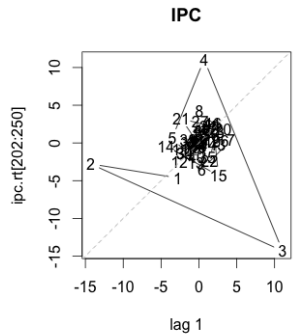

Figure 10 - Lag plot of IPC returns corresponding to Oct 23rd, 1997 to Jan 5th, 1998

IPC returns start to deviate from the main cloud of zero returns at the point marked by 1, corresponding to Oct 23$^{rd}$, 1997 with return (in percentual terms) of -4.64, moving to observation 2 (-2.77), then to the observation 3 (-14.31), to 4 (11.05), and finally going back to the cloud at observation 5 (with return 0.68), after 5 days. Another property of the stock return series that can be inferred from the lag plots presented is that periods of large volatility tend to be triggered by a large negative return.

## Volatility modeling

Volatility clustering can be observed by modeling the conditional variance structure of the time series. The conditional variance of $r_t$, given the past values $r_{t-1}, r_{t-2}, K$, measures the uncertainty in the deviation of $r_t$ from its conditional mean. We have already mentioned how daily returns of stocks are often observed to have larger conditional variance following a period of violent price movement than a relatively stable period. The majority of volatility models in regular use belong to the generalized ARCH (GARCH) family of models. The first of these models was the ARCH model, proposed by Engle (1982), giving way to the GARCH model by (Bollerslev, 1986). Such models are based on using optimal exponential weighting of historical returns to forecast volatility.

Evidence for heteroscedasticity can be shown by performing a McLeod-Li test (plot of the *p*-values of the Box-Ljung statistic applied to squared returns), cf. (McLeod & Li, 1978). Figure 11 shows that the McLeod-Li test statistics are all significant at the 5% significance level and formally shows strong evidence for ARCH in this data.

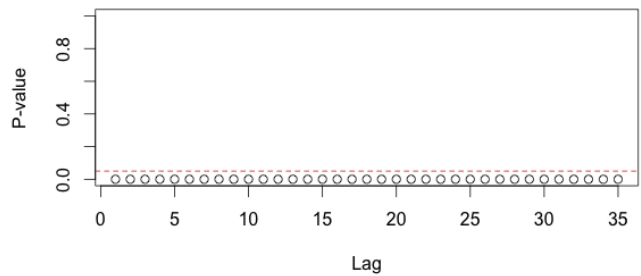

Figure 11 - McLeod-Li test statistics for daily IPC returns

We fit a GARCH(1,1) model to the time series resulting from subtracting the mean from the IPC returns. For more on GARCH models, see, e.g. (Cryer & Chan, 2008). Figure 12 shows the conditional volatility of IPC returns. The full GARCH(1,1) estimation, likelihood and analysis of residuals is beyond the scope of the present work but is available from the authors upon request. A complete study of nonlinearity tests using the GARCH(1,1) model applied to the IPC was recently published by Coronado Ramírez et al. (2012)

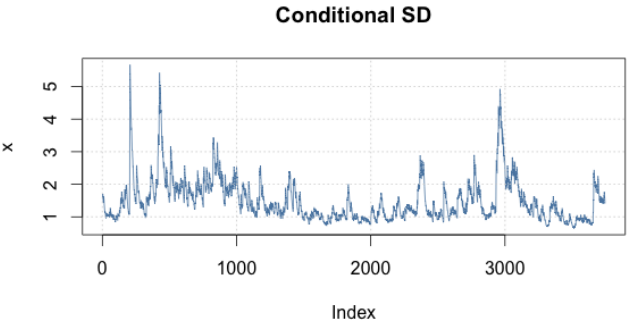

**Figure 12 - Conditional volatility of IPC returns**

Finally, Figure 13 shows the normalized IPC return series with double positive and negative volatility superimposed.

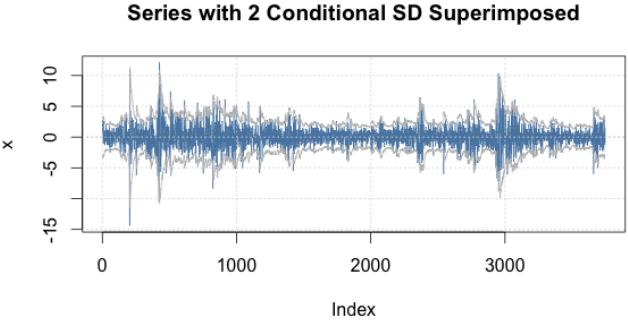

**Figure 13 - Normalized IPC return series with doubled positive and negative volatility superimposed**

**Conclusions**

We have documented some stylized facts exhibited by the return time series of the Mexican Stock Exchange Market (IPC), and compared some of them to other return series, from both developed (USA, UK and Japan) and emerging (Brazil and India) markets. We showed how the probability density function of returns for these indexes is skewed and fat-tailed. The skewness is usually negative, indicating that large market returns are usually negative. However, this was not the case for the IPC and IBOVESPA, implying possible investment opportunities in these emerging markets. Furthermore, fat-tails existed for all markets, with kurtosis far in excess of the corresponding to the normal distribution. Normality of the distribution of the IPC daily returns was rejected using graphical and analytical methods, finding that the kernel density of IPC daily returns was better fitted by a Student t-distribution with four degrees of freedom. However, as the time scale to measure returns is larger (e.g. quarterly), the distribution of IPC returns was better fitted by the normal distribution.

After this, we turned our attention to the autocorrelation function. We find that linear autocorrelations are insignificant after a few lags. However, nonlinear autocorrelations prevailed, which was evidence for the existence of volatility clusters. Such clusters were exhibited using graphical and analytical tools, as well as with the aid of a GARCH(1,1) model.

This study might be thought of as the tip of the iceberg concerning the modeling and forecasting of time series analysis of financial time series. It is helpful to get acquainted with the empirical data before looking for the appropriate models. This is the foundation for our work in progress on nonlinear modeling of financial assets returns, in particular applied to equities. We have focused our future research on threshold autoregressive (TAR) models, both self-exciting and Markov switching.